\documentclass[twocolumn,prd,aps,showpacs,amsmath,amssymb,superscriptaddress,10pt]{revtex4-1}

\usepackage{graphics}

\usepackage{graphicx}

\usepackage{dcolumn}

\usepackage{bm}

\usepackage{mathrsfs}

\usepackage{pstricks}

\usepackage{color}

\usepackage{slashed}

\usepackage{amsmath}

\usepackage{epsfig}

\usepackage{amsfonts}

\usepackage{amssymb}

\def\beq{\begin{equation}}

\def\eeq{\end{equation}}

\def\bea{\begin{eqnarray}}

\def\eea{\end{eqnarray}}

\def\nn{\nonumber}

\def\Re{\textrm{Re}}

\begin{document}

\title{Spontaneous symmetry breaking in replica field theory}

\author{R.~Acosta Diaz}

\email{racosta@cbpf.br}

\affiliation{Centro Brasileiro de Pesquisas F\'{\i}sicas, 22290-180 Rio de Janeiro, RJ, Brazil}

\author{G.~Menezes}

\email{gabrielmenezes@ufrrj.br}

\affiliation{Centro Brasileiro de Pesquisas F\'{\i}sicas, 22290-180 Rio de Janeiro, RJ, Brazil}

\affiliation{Grupo de F\'isica Te\'orica e Matem\'atica F\'isica, Departamento de F\'isica, Universidade Federal Rural do Rio de Janeiro, 23897-000 Serop\'edica, RJ, Brazil}

\author{N.~F.~Svaiter}

\email{nfuxsvai@cbpf.br}

\affiliation{Centro Brasileiro de Pesquisas F\'{\i}sicas, 22290-180 Rio de Janeiro, RJ, Brazil}

\author{C.~A.~D.~Zarro}

\email{carlos.zarro@if.ufrj.br}

\affiliation{Instituto de F\'isica, Universidade Federal do Rio de Janeiro, 21941-972 Rio de Janeiro, RJ, Brazil}


\begin{abstract}

In this paper we discuss a disordered $d$-dimensional Euclidean $\lambda\varphi^{4}$ model.  The dominant contribution to the average free energy of this system is written as a series of the replica 
partition functions of the model. In each replica partition function, using the saddle-point equations and imposing the replica symmetric ansatz, we show the presence of a spontaneous symmetry breaking mechanism in
the disordered model. Moreover, the leading replica partition function must be described by a large-$N$ Euclidean replica field theory. We discuss finite temperature effects considering periodic 
boundary condition in Euclidean time and also using the Landau-Ginzburg approach. In the low temperature regime we prove the existence of $N$ instantons in the model.

\end{abstract}



\maketitle


\section{Introduction}

\label{intro}

One of the most fruitful ideas in classical and quantum field theory has been the concept of spontaneous symmetry breaking. This mechanism is the basis for the construction of renormalizable models of the weak and electromagnetic interactions. On the other hand, in disordered systems the replica symmetry breaking, with its physical consequences, have also been intensively discussed \cite{binder,livro1,livro2,livro3,livro4,0,efetov, parisi1}. Frequently one has to consider averages of extensive quantities, as for example the disorder-dependent free energy~\cite{brout}. Different methods have been developed to compute this quantity. Among them we mention the cavity method \cite{cavity1,cavity2} and the replica method \cite{edwards}, whose several predictions have been confirmed by using other techniques. Concerning the replica approach, a replica symmetry breaking mechanism was introduced by Parisi \cite{RBS1,RBS2,RBS3,RBS4} in order to prevent the emergence of unphysical results (for instance, a negative entropy at low temperatures) which would arise with the assumption of a replica-symmetric solution in a fully-connected mean-field model \cite{SerKir}.  

The basic problems that arise in disordered systems defined in the continuum limit are of two types. First, for a given realization of the disorder the correlation functions $G(x,x')$ depend on both $x$ and $x'$, and not on the difference $x-x'$ as in translational invariant systems. Therefore, since a disordered system is intrinsically inhomogeneous, it is a hard task to perform a perturbative expansion in any model. In addition, in the presence of the disorder field, ground state configurations of the continuous field are defined by a saddle-point equation, where the solutions of such an equation depend on particular configurations of the disorder field. Moreover, it is complicated to implement a perturbative expansion in the situation where there are several local minima in the model. One way to solve both problems is to average the free energy over the disorder field. In this case, one is mainly interested in averaging the disorder-dependent free energy, which amounts to averaging the logarithm of the partition function.

Recently, it was proposed an alternative method to average the disorder-dependent free energy \cite{distributional, distributional2}. In this approach, the dominant contribution to the average free energy is written as a series of the integer moments of the partition function of the model. This method is closely related to the use of spectral zeta-functions for
computing the free energy \cite{seeley,ray,hawking,dowker,casimir,fulling} or the Casimir energy of different systems in quantum field theory. Although different global methods can be used to obtain the Casimir energy of quantum fields, as for example an exponential cut-off or an analytic regularization procedure  \cite{nami1,nami2,nami3}, the spectral zeta-function is powerful and elegant.  
One of the main objectives of the present paper is to discuss, within this framework, the relationship that exists between the spontaneous symmetry breaking mechanism and the replica symmetry ansatz in a disordered scalar model. Such a connection seems to have gone unnoticed so far. 


We are interested in studying a $d$-dimensional Euclidean $\lambda\varphi^{4}$ model in the presence of a disorder field, linearly coupled with a scalar field. The issue of Euclidean fields interacting with delta-correlated disorder has already been investigated in the literature \cite{hamidian1,hamidian2}. In this scenario, recently Aharony and collaborators considered an Euclidean conformal field theory in the presence of disorder \cite{aharony}. In the case where the quantum fluctuations are replaced by the thermodynamic ones, the model discussed here is the continuous version of the random field Ising model in a $d$-dimensional space \cite{mezard1, mezard2, dotsenko, orland, dotsenko3,sherington1,sherington}. For instance, in order to model binary fluids confined in porous media, when the pore surfaces couple differently to the two components of a phase-separating mixture, the random field has been used by the literature \cite{fluids1,fluids2,fluids3}.

The purpose of this paper is to discuss the physical consequences of the adoption of the aforementioned alternative description for evaluation of the average free energy. There are some interesting features brought about by this formalism that we would like to point out. For instance, a connection between spontaneous symmetry breaking mechanism and the structure of the replica space in the disordered model is manifest in the present context. First, in a generic replica partition function, the structure of the replica space is investigated using the saddle-point equations. In order to describe the disorder system, according to the distributional zeta-function method, in each replica partition function we must impose the replica symmetry ansatz, the unique 
solution for the problem of the structure in replica space.
In this scenario, we also show that the system can develop a spontaneous symmetry breaking.  The leading term of this series expansion is a large-$N$ Euclidean replica field theory \cite{LargeN0, LargeN1, LargeN2}.

Next, we discuss finite temperature effects in the disordered model. Finite size effects in quantum field theory \cite{finitez1,finitez2,finitez3,finitez4,khanna}, critical phenomena \cite{danchev} and classical random systems \cite{s1,s2,ricci1,ricci2} are areas of intense activity in the last years. 
In the Ref. \cite{periodic}, it was discussed finite-size effects in the disordered $\lambda\varphi^{4}$ model, applying the standard replica method in the one-loop approximation and also using a gap equation \cite{jackiw,ananos,gap}. Questions concerning the nature of the phase transition in the continuous version of the random field Ising model in a $d$-dimensional space can be analyzed, following the Landau-Ginzburg approach. 
Using a mean-field description for phase transitions, we show that for high temperatures, in the large-$N$ approximation, the symmetry $[\mathbb{Z}_{2}\times\mathbb{Z}_{2}\cdots\times\mathbb{Z}_{2}]$ is realized. For low temperatures, taking only the leading order term of the series that represent the average free energy, i.e., the large-$N$ approximation, the symmetry $[\mathbb{Z}_{2}\times\mathbb{Z}_{2}\cdots\times\mathbb{Z}_{2}]$ is broken. In order to go beyond the tree-level approximation, in the replica field theory, we also consider periodic boundary conditions in Euclidean time. We discuss the dependence of the renormalized mass on the radius of the compactified dimension in a scenario of spontaneous symmetry breaking in the one-loop approximation. We prove that there is a critical temperature where the renormalized mass vanishes. 

Another interesting issue concerns the presence of instantons in the model at very low temperature. We study the dominant replica partition function using a representation closely related to the strong-coupling expansion in field theory investigated in Refs. \cite{SKD,RMDHS,CBFC,sce}. See also the linked-cluster expansion \cite{lusher1,lusher2,lusher3,reisz1,reisz2}. The main difference is that in our case, instead of an independent-value action we have a functional differential operator connecting replica fields acting on a modified replica partition function. The first term in this perturbative expansion is the diluted instanton approximation. We show that for $\sigma N>m_{0}^{2}>-3\sigma N$, where $\sigma$ is a parameter that characterizes the strength of the disorder, one finds the existence of $N$ complex instantons in the model; furthermore, for $m_{0}^{2}<-3\sigma N$ such instantons are real \cite{dotsenko4,dotsenko2,Maxin,Allan}.

The organization of this paper is the following. In Sec. II, using the distributional zeta-function method we discuss each replica field theory of the disordered $\lambda\varphi^{4}$ model. In Sec. III, in a generic replica partition function we discuss the structure of the replica space using the saddle-point equations of the model. In Sec. IV, we discuss temperature effects in the replica field theory. In Sec. V we demonstrate the emergence of $N$ instantons in the model. Conclusions are given in Sec. VI. We use units such that $\hbar=c=k_{B}=1$.

\vspace{.05cm}


\section{From the disordered model to the replica field theory}

\label{model}

The aim of this section is to obtain replica field theories from an Euclidean scalar field theory in the presence of a disorder field. In the functional integral formulation of field theory there are two kinds of random variables. The first ones are the Euclidean fields. These fields describes generalized Euclidean processes with zero mean and covariance defined in terms of gradients. There are also variables, the disorder fields, with the absence of any differential operator. For such fields the two-point correlation function is not defined in terms of gradients. These are the non-propagating degrees of freedom of the theory.

Let us assume an Euclidean $d$-dimensional $\lambda\varphi^{4}$ model in the presence of a disorder field, where the disordered functional integral $Z(h)$ is  defined by 

\begin{equation}
Z(h)=\int [d\varphi]\,\, \exp\left(-S+ \int d^{d}x\,h(x)\varphi(x)\right). \label{8}
\end{equation}

\noindent In the above equation $S=S_{0}+S_{I}$  is the Euclidean-invariant action functional of the real scalar field where $S_{0}(\varphi)$, given by

\begin{equation}
S_{0}(\varphi)=\int d^{d}x\, \left(\frac{1}{2}(\partial\varphi)^{2}+\frac{1}{2}m_{0}^{2}\,\varphi^{2}(x)\right), \label{9}
\end{equation}

\noindent is the free field Euclidean action and $S_{I}(\varphi)$, defined by

\begin{equation}
S_{I}(\varphi)= \int d^{d}x\,\frac{\lambda_{0}}{4!} \,\varphi^{4}(x),\label{10}
\end{equation}

\noindent is the self-interacting non-Gaussian contribution. In Eq. (\ref{8}), $[d\varphi]$ is a formal Lebesgue measure, given by $[d\varphi]=\prod_{x} d\varphi(x)$. Usually, the quantities $\lambda_{0}$ and $m_{0}^{2}$ are, respectively, the bare coupling constant and the mass squared of the model. Finally, $h(x)$ is a disorder field of the Euclidean field theory. For simplicity we are assuming  a linear coupling between the Euclidean scalar field and the disorder field. In order to construct a probability measure a normalization factor is introduced.  We must have $Z(h)|_{h\rightarrow 0}=1$. For simplicity the normalization factor is absorbed  in the formal Lebesgue measure. We impose to the scalar field for example Dirichlet boundary conditions, $\varphi(x)\rightarrow 0$ as $|x|\rightarrow \infty$. We want to stress that our starting point is the semiclassical (tree) approximation. In this case $\lambda_{0}$ and $m_{0}$ are  renormalized quantities. Consequently, our discussion will be in the tree-level approximation until Section IV.

There are two different ways to eliminate the disorder field. For a given probability distribution $P(h)$ of the disorder, one average the disordered functional integral $Z(h)$ and take the logarithm of this quantity. Then we define the annealed free energy $F_{a}$ as

\begin{equation}
F_{a}=-\ln\biggl(\int\,[dh]P(h)\,Z(h)\biggr).\label{afe}
\end{equation}

Here we wish to obtain a different free energy, that is called the quenched free energy in the literature. For a given probability distribution of the disorder, one is mainly interested in averaging the logarithm of the disordered functional integral $Z(h)$. Relying on the similarity upon statistical mechanics we call it the disorder-dependent free energy $F(h)$.   
It reads

\vspace{-2mm}

\begin{eqnarray}
\hspace{-1.2mm}
F(h)=&&-\ln\, \int[d\varphi]  \exp\left[-\int d^{d}x\, \biggl(\frac{1}{2}\varphi(x)\Bigl(-\Delta\ +m_{0}^{2}\Bigr)\varphi(x) \right.
\nonumber\\
\hspace{-1.5mm}
&&+\, \left. \frac{\lambda_{0}}{4!}\varphi^{4}(x)-h(x)\varphi(x)\biggr)\right],\label{fe}
\end{eqnarray}

\noindent where the symbol $\Delta$ denotes the Laplacian in $\mathbb{R}^{d}$. The average free energy $F_{q}$ is defined as

\begin{equation}
F_{q}=-\int\,[dh]P(h)\ln Z(h), \label{sa2}
\end{equation}

\noindent where $[dh]=\prod_{x} dh(x)$ is again a formal Lebesgue measure. To justify the Eq. (\ref{sa2}), let us first assume a compact domain. Suppose that we divide the domain in subsystems. We consider each subsystem representing a realization of the disorder field and that the coupling between the subsystems is negligible. The value of any extensive variable for the whole system is equal to the average of the values of this quantity over the subsystems. From the extensive property of the free energy we get the self-averaging property. Next, using the fact that the self-averages of the free energy hold when the domain is non-compact, we justify the averaging of the logarithm of the partition function.

Coming back to the problem, we assume that the disorder field $h(x)$ is described by a Gaussian distribution, i.e., the probability distribution of the disorder, is written as $[dh]\,P(h)$, where

\begin{equation}
P(h)=p\,\exp\Biggl(-\frac{1}{2\,\sigma}\int\,d^{d}x(h(x))^{2}\Biggr).\label{dis2}
\end{equation}

\noindent The quantity $\sigma$ is a positive parameter associated with the disorder and $p$ is
a normalization constant. In this case we have a delta correlated disorder field, i.e., $\mathbb{E}[{h(x)h(y)}]=\sigma\delta^{d}(x-y)$.

An established technique for computing the average free energy is the replica method. 
This consists in the following steps. First, one constructs the (integer) $k$-th power of the partition function $Z^{k}(h)$. Second, the expected value of the partition function's $k$-th power $\mathbb{E}\,[ Z^k(h)]$ is computed by integrating over the disorder field on the new model. Finally, the average free energy is obtained using the formula 

\[\mathbb{E}\,[{\,\ln Z(h)}]=\lim_{k\rightarrow 0}\frac{\partial}{\partial k}\mathbb{E}\,[ Z^k(h)].\]

\noindent The average value of the free energy in the presence of the disorder is then obtained taking the limit $k\rightarrow 0$.

An alternative approach to compute the average free energy of disordered systems was presented in Refs. \cite{distributional,distributional2}. We call it the distributional zeta-function method. An attractive characteristic of such a method is that one can find an analytic expression for the free energy unlike the standard replica method as it involves derivatives of the integer moments of the partition function. Observing that if  $(X,\mathcal{A},\mu)$ is a measure space and $f:X\to(0,\infty)$ is measurable, a generalized $\zeta$-function is defined

\begin{equation}
\zeta_{\,\mu,f}(s)=\int_X f(x)^{-s}\, d\mu(x),
\end{equation}

\noindent for those $s\in\mathbb{C}$ such that  $f^{-s}\in L^1(\mu)$, where in the above integral $f^{-s}=\exp(-s\log(f))$ is obtained using the principal branch of the logarithm. This formalism contains some well-know examples of zeta-functions for $f(x)=x$: the Riemann zeta-function \cite{riem,riem2} is obtained if $X=\mathbb{N}$ and $\mu$ is the counting measure; however, if $\mu$ counts only the prime numbers, we get the prime zeta-function \cite{landau,fro}; if $X=\mathbb {R}$ and $\mu$ counts the eigenvalues of an elliptic operator, with their respective multiplicity, the spectral zeta-function is obtained. Further extending this formalism for the case where $f(h)=Z(h)$ and $d\mu(h)=[dh]P(h)$ leads to the definition of the distributional zeta-function $\Phi(s)$ as

\begin{equation}
\Phi(s)=\int [dh]P(h)\frac{1}{Z(h)^{s}},\label{pro1}
\vspace{.2cm}
\end{equation}

\noindent for $s\in \mathbb{C}$, this function being defined in the region where the above integral converges. Note that the free energy of the system with annealed disorder is given by

\begin{equation}
F_{a}=-\ln\,\Phi(s)|_{s=-1}.
\end{equation}

Following the usual steps of the spectral zeta function, the average free energy $F_{q}$ can be written as

\begin{equation}
F_q=(d/ds)\Phi(s)|_{s=0^{+}}, \,\,\,\,\,\,\,\,\,\, \Re(s) \geq 0, 
\end{equation}

\noindent where $\Phi(s)$ is well defined. Using analytic tools, the average free energy can be represented as

\begin{equation}
F_q=\sum_{k=1}^\infty \frac{(-1)^{k}a^{k}}{k!k}\,\mathbb{E}\,[{Z^{\,k}}]+\bigl(\ln(a)+\gamma\bigr)-R(a)
\label{m23e}
\end{equation}

\noindent where

\begin{equation}
\left|R(a)\right|\leq \dfrac{1}{Z(0)a}\exp\big(-Z(0)a\big),
\end{equation}

\noindent with $a$ being an arbitrary dimensionless constant.

In the Eq.~(\ref{m23e}), the free energy is independent of $a$. Since we are not able to estimate the contribution of $R(a)$ to the free energy, an approximation is necessary. The contribution of $R(a)$ to the free energy can be made as small as desired, taking $a$ large enough such that $a \gg 1/Z(0)$. As we will see, this system must be described by a large-$N$ Euclidean replica field theory where the dimensionless parameter $a$ can be absorbed in the formal Lebesgue measure.
With respect to this fact, a remark is in order. All the local quantities obtained from the replica partition function are independent of $a$, as for example, the two-point correlation function in the field theory formulation for a directed polymer and an interface in a quenched random potential \cite{polymer}.

From Eq. (\ref{m23e}), we have to compute the replica partition function $\mathbb{E}\,[{Z^{\,k}}]$. First, it is easy to show that $Z^{\,k}$  is given by

\begin{equation}
(Z(h))^{k}=\int\,\prod_{i=1}^{k}[d\varphi_{i}]\,\exp\biggl(-\sum_{i=1}^{k}S(\varphi_{i},h)\biggr).
\label{u1}
\end{equation}

\noindent Moreover, using the probability distribution of the disorder defined by Eq. (\ref{dis2}), after integrating over the disorder we get that a generic replica partition function can be written as

\begin{equation}
\mathbb{E}\,[{Z^{\,k}}]=\int\,\prod_{i=1}^{k}\left[d\varphi_{i}\right]\,\exp\biggl(-S_{\textrm{eff}}\left(\varphi_{i}\right)\biggr),\label{aa11}
\end{equation}

\noindent where the effective action $S_{\textrm{eff}}\left(\varphi_{i}\right)$ is given by

\bea
S_{\textrm{eff}}\left(\varphi_{i}\right)&=&\frac{1}{2}\sum_{i,j=1}^{k} \int d^{d}x\int d^{d}y\,\,\varphi_{i}(x)D_{ij}(x-y)\varphi_{j}(y)\nn\\
&+&\,\frac{\lambda_{0}}{4!}\sum_{i=1}^{k}\int\,d^{d}x\, \varphi_{i}^{4}(x).\label{aa12}
\eea

\noindent In the above equation we have that $D_{ij}(x-y)=D_{ij}(m_{0},\sigma;x-y)$ where

\begin{equation}
D_{ij}(m_{0},\sigma;x-y)=\biggl(\delta_{ij}(-\Delta+m_{0}^{2})-\sigma\biggr)\delta^{d}(x-y).
\label{aa124}
\end{equation}

\noindent The Eqs. (\ref{aa11}), (\ref{aa12}) and (\ref{aa124}) are similar to an Euclidean field theory for $k$ interacting replica fields. Being more precise, the Eq. (\ref{aa11}) with an external source is the generating functional of the correlation functions of the model. Using a statistical mechanics language, we call it a replica partition function.

Let us analyze these results in the momentum space. After a Fourier transform,  we obtain

\bea
S_{\textrm{eff}}\left(\varphi_{i}\right)&=&\frac{1}{2}\sum_{i,j=1}^{k}\int\,\frac{d^{d}p}{(2\pi)^{d}}\,\,\varphi_{i}(p)\bigl[G_{0}\bigr]_{ij}^{-1}\varphi_{j}(-p)\nn\\
&+&\,\frac{\lambda_{0}}{4!}\sum_{i=1}^{k}\varphi_{i}^{4},\label{aa13}
\eea

\noindent where in the quadratic part of $S_{\textrm{eff}}\left(\varphi_{i}\right)$, the quantity $\bigl[G_{0}\bigr]_{ij}^{-1}$  is defined as

\begin{equation}
\bigl[G_{0}\bigr]_{ij}^{-1}(p)=(p^{2}+m_{0}^{2})\delta_{ij}-\sigma,\label{aa14}
\end{equation}

\noindent which can be inverted; hence

\begin{equation}
\hspace{-1.2mm}
\bigl[G_{0}\bigr]_{ij}(p)=\frac{\delta_{ij}}{(p^{2}+m_{0}^{2})}+\frac{\sigma}{(p^{2}+m_{0}^{2})(p^{2}+m_{0}^{2}-k\sigma)}.
\label{aa18}
\end{equation}

\noindent This is the two-point correlation function of the replica field theory in the tree-level approximation.


\section{Spontaneous symmetry breaking in replica field theory} \label{replica}

The aim of this section is to show the presence of a spontaneous symmetry breaking mechanism in the disordered model. 
In the presence of the disorder field linearly coupled with the scalar field, ground state configurations of the field $\varphi(x)$ are defined by a saddle-point equation, where
the solutions of such equation depends on particular configurations of the disorder fields. The saddle-point equation of the disordered model reads

\begin{equation}
\Bigl(-\Delta\,+m_{0}^{2}\Bigr)\varphi_{_{h}}(x)+\frac{\lambda_{0}}{3!}\varphi^{3}_{_{h}}(x)=h(x),\label{sp2}
\end{equation}

\noindent where $\varphi_{_{h}}$ means the field $\varphi$ defined for a particular configuration of the disorder. The situation with several local minima in the model precludes the realization of a perturbative expansion in a straightforward way. After integrating out the disorder field in a generic replica partition function, $\mathbb{E}\,[{Z^{\,k}}]$, the saddle-point equations reads

\begin{equation}
\Bigl(-\Delta\,+m_{0}^{2}\Bigr)\varphi_{i}(x)+\frac{\lambda_{0}}{3!}\varphi_{i}^{3}(x)=\sigma\sum_{j=1}^{k}\varphi_{j}(x).\label{sp}
\end{equation}

\noindent According to the distributional zeta-function method, the average free energy is written as a series of the integer moments of the partition function of the model. Therefore, the only choice in each replica partition function is the replica symmetric ansatz, i.e., all replica fields must be equal in each replica partition function, $\varphi_{i}(x)=\varphi_{j}(x)$. This choice then implies that the saddle-point equations read

\begin{equation}
\Bigl(-\Delta\,+m_{0}^{2}-k\sigma\Bigr)\varphi_{i}(x) +\frac{\lambda_{0}}{3!}\varphi_{i}^{3}(x)=0.\label{sp}
\end{equation}

\noindent In principle, observe that within this approach one must take into account all replica partition functions contributing to the average free energy, i.e., all values of $k$ must be considered. In the following, in order to proceed, we are assuming $m_{0}^{2}>0$. Let us define a critical $k_{c}$ such that $k_{c}=\left\lfloor m_{0}^{2}/\sigma\right\rfloor$ where $\left\lfloor x\right\rfloor$ means the integer part of $x$. For $m_{0}^{2}\geq\sigma$, in a generic replica partition function, $m_{0}^{2}-k\sigma\,\geq 0$ is satisfied as $k\leq k_{c}$. In such a case, each replica field fluctuates around the zero value, the stable equilibrium state. One must notice a interesting fact here, the effective mass of the replica fields in different replica partition functions are not equal. This situation is quite different when the contribution to the average free energy comes from the replica partition functions where $k>k_{c}$. From Eq. (\ref{sp}), all of these replica fields fluctuate around the zero value which is not an equilibrium state anymore.  In the framework of field operators, this means that the vacuum expectation value of such fields do not vanish. This is exactly the scenario that a spontaneous symmetry breaking emerges.

Before continuing, we would like to summarize the main differences between the consequences of our formalism and the standard replica method. In the standard replica method, in the replica partition function, we must take the limit $k\rightarrow 0$. After choosing the replica symmetric ansatz, the saddle-point equation reduces to the standard model without disorder. In our formalism, for each replica field theory, investigating the saddle-point equations and imposing the replica symmetric ansatz we obtain a critical $k_{c}$. We can now ask what assumptions we must use to circumvent the above mentioned problem.

The point that we wish to stress is that due to replica fields for replica partition functions such that $k>k_{c}$, a spontaneous symmetry breaking mechanism occurs. To proceed, let us investigate some choices in the replica space. A interesting question is whether there are 
different choices for replica symmetry breaking. Consider a generic term of the series given by Eq. (\ref{m23e}) with replica partition function given by $\mathbb{E}\,[{Z^{\,l}}]$. One choice in 
the structure of the fields in each replica partition function is given by

\begin{equation}
\begin{cases}
\varphi_{i}^{(l)}(x)=\varphi(x) \,\,\,\hfill\hbox{for $l=1,2,...,k_{c}$}\\
\varphi_{i}^{(l)}(x)=0 \quad \,\,\,\,\,\hbox{for $l>k_{c}$},
\end{cases} \label{RSB1}
\end{equation}

\noindent where for the sake of simplicity we still employ the same notation for the field. However, the effect of this choice may represent a very constraining truncation for the series representation of the average free energy, given by Eq. (\ref{m23e}). Indeed, as discussed previously, this choice in replica space is not consistent with the distributional zeta-function method. In order to take into account more terms in this series, we consider $N>k_{c}$, where $m_{0}^{2}-k\sigma\,<0$, for $N>k>k_{c}$. To proceed, we must study in each replica partition function the vacuum structure that emerges in our scenario. In this situation we must consider the following structure of the replica space

\begin{align}
\hspace{-2.2mm}
\begin{cases}
\varphi_{i}^{(l)}(x)=\varphi(x) \,\,\,\,\,\,\,\quad \hbox{for $l=1,\ldots,k_{c}$ and $i=1,\ldots,l$} \\
\varphi_{i}^{(l)}(x)=\phi(x)+v  \,\,\, \hbox{for $l=k_{c}+1,\ldots,N$ and $i=1,\ldots,l$} \\
\varphi_{i}^{(l)}(x)=0 \,\,\,\,\,\,\,\,\,\,\qquad\hbox{for $l>N$,}
\end{cases} 
\label{RSB2}
\end{align}

\noindent where

\begin{equation}
v=\left(\frac{6(\sigma N-m_{0}^{2})}{\lambda_{0}}\right)^{1/2}.
\end{equation}

In terms of these new shifted fields, we get a positive mass squared with new self-interaction vertices $\phi^{3}$ and $\phi^{4}$. There is a spontaneous symmetry breaking for a finite $N$. We are interested in the case with large-$N$, which will be discussed in the following sections. This structure in replica space, defined by Eq. (\ref{RSB2}), also 
with the large-$N$ limit is quite natural and it is the only choice compatible with the method developed in Refs. \cite{distributional,distributional2}. In conclusion, our arguments stated here show the uniqueness of the solution for the problem of the structure in replica space. Notice that all replica fields are the same in each replica partition function. This is not true anymore for different replica partition functions. Being more precise, in the scenario constructed by the replica method, the breaking of replica symmetry in a unique replica partition function occurs by choosing different replica fields. In  the distributional zeta-function method, \emph{a priori}, all replica fields are the same in each replica partition function. Since the replica fields of different replica partition function are different, we also call it replica symmetry breaking. 

In the following, we are using the structure of  replica space given by Eq. (\ref{RSB2}). With this choice, the dominant contribution to the average free energy can be written as

\begin{equation}
F_q(a)=\sum_{k=1}^{N} \frac{(-1)^{k}a^{k}}{k!k}\,\mathbb{E}\,[{Z^{\,k}}]. 
\label{KMenergy}
\end{equation}

\noindent Notice that this series representation has two kinds of replica partition functions. For $k\leq k_{c}$, $\mathbb{E}\,[{Z^{\,k}}]$ is given by Eqs. (\ref{aa11}), (\ref{aa12}) and 
(\ref{aa124}). For $k_{c}<k\leq N$, the replica partition function $\mathbb{E}\,[{Z^{\,k}}]$ is

\begin{equation}
\mathbb{E}\,[{Z^{\,k}}]=\int\,\prod_{j=1}^{k}\left[d\phi_{j}\right]\,\exp\biggl(-S_{\textrm{eff}}\left(\phi_{j}\right)\biggr), \label{ZKM}
\end{equation}

\noindent where $S_{\textrm{eff}}\left(\phi_{j}\right)$ is given by

\begin{eqnarray}
\hspace{-7.5mm}
S_{\textrm{eff}}\left(\phi_{i}\right)&=&\frac{1}{2}\sum_{i,j=1}^{k}\int d^{d}x\int d^{d}y\,\,\phi_{i}(x)C_{ij}(x-y)\phi_{j}(y)
\nonumber\\
&+&\frac{\lambda_{0} v}{3!}\sum_{i=1}^{k}\int d^{d}x \phi_{i}^{3}(x)
+\frac{\lambda_{0}}{4!}\sum_{i=1}^{k}\int d^{d}x \phi_{i}^{4}(x),
\label{Kaction}
\end{eqnarray}

\noindent and the operator $C_{ij}(x-y)$ is

\begin{equation}\label{Cij}
C_{ij}(x-y)=\biggl[\bigl(-\Delta+3\sigma N-2m_{0}^{2}\bigr)\delta_{ij}-\sigma\biggr]\delta^{d}(x-y).
\end{equation}

\noindent The Eq.~(\ref{Kaction}) is the first important result of the paper, i.e., the spontaneous symmetry breaking in the disorder scenario. As in the case without spontaneous symmetry breaking, performing a Fourier transform from the quadratic part, in the Eq. (\ref{Kaction}), we can again identify the inverse of the two-point correlation function of the replica field theory which is now given by

\begin{equation}
[G_{0}]_{ij}^{-1}=\bigl(p^{2}+3\sigma N-2m_{0}^{2}\bigr)\delta_{ij}-\sigma.
\end{equation}

\noindent Using the projector operators we can write the correlation function $[G_{0}]_{ij}$ as

\begin{align}
\hspace{-1mm}
\bigl[G_{0}\bigr]_{ij}(p)&=\frac{\delta_{ij}}{(p^{2}+3\sigma N-2m_{0}^{2})}\nonumber\\
&+\frac{\sigma}{(p^{2}+3\sigma N-2m_{0}^{2})(p^{2}+\sigma(3N-k)-2m_{0}^{2})}.
\label{Kcorr}
\end{align}

We shall now examine the presence of Goldstone bosons in the model. The main difference between the usual situation in the literature and scenario discussed by us is that Goldstone bosons appear when there is a breaking of a continuous symmetry. There are no Goldstone bosons in the model, since we are breaking a discrete symmetry. This issue will be clarified in Sec. \ref{landauginzburg}. As we discussed before, for $a$ large enough, the leading term in the series representation defined by Eq. (\ref{KMenergy}) is given by $k=N$. In this situation the replica partition function, $\mathbb{E}\left[Z^{N}\right]$, for $m_{0}^{2}\geq\sigma N$, all the replica fields are oscillating aroung the trivial vacuum. For $m_{0}^{2}<\sigma N$, all the replica fields now oscillate around the non-trivial vacuum. In this case, the replica partition function reads 


%

\begin{equation}
\mathbb{E}\,[{Z^{N}}]=\int\,\prod_{i=1}^{N}[d\phi_{i}]\,\exp\biggl(-S_{\textrm{eff}}(\phi_{i})\biggr),
\label{NZ}
\end{equation}

\noindent where the effective action $S_{\textrm{eff}}(\phi_{i})$ is given by

\begin{align}
S_{\textrm{eff}}(\phi_{i})&=\frac{1}{2}\sum_{i,j=1}^{N}\int d^{d}x\int d^{d}y\,\,\phi_{i}(x)C_{ij}(x-y)\phi_{j}(y) \nonumber \\
&+\frac{\lambda_{0} v}{3!}\sum_{i=1}^{N}\int d^{d}x\, \phi_{i}^{3}(x) +\frac{\lambda_{0}}{4!}\sum_{i=1}^{N}\int\,d^{d}x\, \phi_{i}^{4}(x),\label{KactionN}
\end{align}

\noindent and $C_{ij}(x-y)$ is given by Eq. (\ref{Cij}).


Let us summarize our results. The leading contribution for the free energy consists in a series in which all the replica partition functions contribute. The subtle issue here is that as we perform an expansion in the integer moments of the partition function, we choose the structure in the replica space with the most symmetric case, namely all replica fields are the same in each replica partition function. All of the above discussion lead us to the large-$N$ scenario in replica field theory. Notice that instead of having one 't Hooft coupling, which means that $g_{0}=\lambda_{0} N$ is finite although $N\rightarrow \infty$ and $\lambda_{0}\rightarrow 0$, we also have another 't Hooft coupling, $f_{0}=\sigma N$ which is finite although $N\rightarrow\infty$ and $\sigma\rightarrow 0$ (weak disorder). Here, we are mainly interested in the situation where the disorder is weak. In this context we have just established a path to clarify the relationship between two hitherto unconnected results. It is known that for $d>6$, the critical region in the random field Ising model can be described using the mean-field exponents \cite{Grinstein}. In turn, in the $O(N)$ symmetric field theory of any real scalar fields with interaction $\lambda_{0}\left(\varphi^{2}_{i}\varphi^{2}_{i}\right)^{2}/4!$, the $1/N$ expansion for $d>6$ is not useful \cite{parisi1}. Hence the $1/N$ expansion is efficient when disorder affects the critical region in the random field Ising model in a non-trivial way. We interpret this connection as a consequence of approaching quenched disorder in a large-$N$ scenario in replica field theory. In any case, despite the above remark we assert that all calculations can be carried out irrespective of the space dimensions; in particular for $d < 6$ one may resort to a large-$N$ expansion.


\section{Temperature effects in the replica field theory}  \label{temperature}

The aim of this section is to discuss temperature effects in the replica field theory defined by Eqs. (\ref{NZ}) and (\ref{KactionN}). 
As we discussed before, when the quantum fluctuations are replaced by the thermodynamic ones, the model studied is the continuous version of the random field Ising model in a $d$-dimensional space. In order to describe the phase transition in this model we follow the Landau-Ginzburg phenomenological approach where, for a system without disorder, the mass squared depends on the reduced temperature, defined by $t=(T-T_{c})/T_{c}$, where $T$ is the temperature and $T_{c}$ is the critical temperature of the system. On the other hand, in order to go beyond the tree-level approximation with quantum fluctuations one must take into account loop corrections. With this respect we depart from the Landau-Ginzburg formalism and assume instead that the fields in the replica field theory satisfy periodic boundary condition in Euclidean time. 

\vspace{1mm}

\subsection{Landau-Ginzburg approach in replica field theory}\label{landauginzburg}

The model considered in this work is also the continuous version of the random field Ising model in a $d$-dimensional space, where the dependence from the temperature is concentrated in $m_{0}^{2}$.  In the following, we continue to use the semiclassical (tree) approximation. 
For a system without disorder at sufficiently high temperatures there is no spontaneous symmetry breaking, where the system presents a $\mathbb{Z}_{2}$ symmetry. On the other hand, in the low temperature regime ($T<T_{c}$), we have a spontaneous symmetry breaking, $i.e.$, the $\mathbb{Z}_{2}$-symmetry is broken.

In this disordered system, this situation is more involved, since the average free energy is written as a series defined by Eq. (\ref{KMenergy}). Inspired in the above situation, we will assume that $m_{0}^{2}$ depends on the temperature, it is not positive definite and is a monotonically increasing function on the temperature. For simplicity, let us assume that the disorder is weak and fixed. Before taking the large-$N$ limit, one has three interesting cases, with two  temperatures, $T^{(1)}_{c}$ and $T^{(2)}_{c}$.

\begin{itemize}

\item[\bf I.] For temperatures such that $m_{0}^{2}\geq\sigma N$, all the replica fields in the replica partition functions in Eq. (\ref{KMenergy}) oscillate around the trivial vacuum $\varphi=0$. In this case, for a very large $a$, the average free energy is written as

\bea
F_q(a)&=&\sum_{k=1}^{N} \frac{(-1)^{k}a^{k}}{k!k}\,\mathbb{E}^{(1)}\,[{Z^{\,k}}],  
\label{LGEnergyCase1}
\eea

\noindent where the replica partition functions $\mathbb{E}^{(1)}[{Z^{\,k}}]$ are

\begin{equation} \label{E1(Zk)}
\mathbb{E}^{(1)}[{Z^{\,k}}]=\int\,\prod_{i=1}^{k}\left[d\varphi_{i}\right]\,\exp\biggl(-S^{(1)}_{\textrm{eff}}\left(\varphi_{i}\right)\biggr).
\end{equation}

\noindent The effective action $S^{(1)}_{\textrm{eff}}\left(\varphi_{i}\right)$ is given by

\bea
S_{\textrm{eff}}^{(1)}(\varphi_{i})&&=\int d^{\,d}x\Biggl[\sum_{i=1}^{k}\biggl(\frac{1}{2}\varphi_{i}(x)\bigl(-\Delta+m_{0}^{2}\bigr)\varphi_{i}(x)\nn\\
&&+\frac{g_{0}}{4!N}\varphi_{i}^{4}(x)\biggr)-\frac{f_{0}}{2N}\sum_{i,j=1}^{k}\varphi_{i}(x)\varphi_{j}(x)\Biggr].
\label{Seff(1)}
\eea

\noindent In the large-$N$ limit, such that $a \gg N$, the leading term of the series of the average free energy is given by the replica partition function with $N$ fields $\varphi_{i}$. Hence, we have the symmetry $[\mathbb{Z}_{2}\times\mathbb{Z}_{2}\cdots\times\mathbb{Z}_{2}]$ for $N$ replica fields. The temperature $T^{(1)}_{c}$ occurs when $m_{0}^{2}=N\sigma$. Below this temperature, $[\mathbb{Z}_{2}\times\mathbb{Z}_{2}\cdots\times\mathbb{Z}_{2}]$ symmetry is broken.

\item[\bf II.] For $\sigma N>m_{0}^{2}\geq\sigma$, the temperature decreases. Before taking the large-$N$ limit, all the replica fields of some replica partition functions oscillate around the non-trivial vacuum, and all the replica fields of the remaining replica partition functions oscillate around $\phi=0$.  Defining $k_{c}(T)=\left\lfloor m_{0}^{2}(T)/\sigma\right\rfloor$, we can write the series representation of the average free energy in the Landau-Ginzburg approach as   

\bea
F_q(a)&=&\sum_{k=1}^{k_{c}(T)} \frac{(-1)^{k}a^{k}}{k!k}\,\mathbb{E}^{(1)}\,[{Z^{\,k}}]\nn\\
             &+&\sum_{k=k_{c}(T)+1}^{N} \frac{(-1)^{k}a^{k}}{k!k}\,\mathbb{E}^{(2)}\,[{Z^{\,k}}],  
\label{LGEnergy}
\eea

\noindent where $\mathbb{E}^{(1)}[{Z^{\,k}}]$ is given by Eq. (\ref{E1(Zk)}) and 

\begin{equation} \label{E2(Zk)}
\mathbb{E}^{(2)}[{Z^{\,k}}]=\int\,\prod_{j=1}^{k}\left[d\phi_{j}\right]\,\exp\biggl(-S^{(2)}_{\textrm{eff}}\left(\phi_{j}\right)\biggr).
\end{equation}

\noindent The effective action $S_{\textrm{eff}}^{(2)}(\phi_{i})$ is written as

\bea
\hspace{2mm}
S_{\textrm{eff}}^{(2)}(\phi_{i})&&=\int d^{\,d}x\Biggl[\sum_{i=1}^{k}\biggl(\frac{1}{2}\phi_{i}(x)\bigl(-\Delta+3f_{0}-2m_{0}^{2}\bigr)\phi_{i}(x)\nn\\
&&\,+\Bigl(\frac{f_{0}g_{0}}{3!N}\Bigr)^{\frac{1}{2}}\left(1-\frac{m_{0}^{2}}{f_{0}}\right)^{\frac{1}{2}}\phi_{i}^{3}(x)+\frac{g_{0}}{4!N}\phi_{i}^{4}(x)\biggr)\nn\\
&&-\frac{f_{0}}{2N}\sum_{i,j=1}^{k}\phi_{i}(x)\phi_{j}(x)\Biggr].
\label{Seff(2)}
\eea

Therefore, in this region one has two types of replica partition functions in the series representation of the average free energy. In the large-$N$ approximation, using again that $a\gg N$, the average free energy is described by a unique replica partition function with all replica fields oscillating around  the non-trivial vacuum.

\item[\bf III.] For $m_{0}^{2}<\sigma$, all the replica fields in replica partition functions are oscillating around the non-trivial vacuum. The temperature $T^{(2)}_{c}$ is given by $m_{0}^{2}=\sigma$. The average free energy describing this case is given by

\bea
F_q(a)&=&\sum_{k=1}^{N} \frac{(-1)^{k}a^{k}}{k!k}\,\mathbb{E}^{(2)}\,[{Z^{\,k}}], \label{LGEnergyCase3}
\eea

\noindent where $\mathbb{E}^{(2)}[{Z^{\,k}}]$ is given by Eq. (\ref{E2(Zk)}).

For $a\gg N$,  a very large $N$ limit consists in taking the leading term of the series, which  is given by a unique replica partition function with $N$ replica fields $\phi_{i}$. This situation is equivalent to the $\mathbb{Z}_{2}$-broken symmetry for a system without disorder. The symmetry $[\mathbb{Z}_{2}\times\mathbb{Z}_{2}\cdots\times\mathbb{Z}_{2}]$ for $N$ replica fields remains broken. 

\end{itemize}

In summary, in the disordered system, before taking the large-$N$ approximation, there are two temperatures, $T^{(1)}_{c}$ and $T^{(2)}_{c}$. Above $T^{(1)}_{c}$ the average free energy is written by a series of replica partition functions where in all of them the replica fields are oscillating around the trivial vacuum. Below $T^{(1)}_{c}$ and above $T^{(2)}_{c}$ the average free energy is defined by two kinds of replica partition functions with replica fields $\varphi_{i}$ and $\phi_{i}$ respectively. In the large-$N$ limit, as only the leading term is considered, one has that all the replica fields of this leading replica partition function are oscillating around the non-trivial vacuum. Below $T^{(2)}_{c}$ all the replicas partition functions that define the average free energy are composed by $\phi_{i}$ fields. In the large-$N$ regime, as one is forced to consider the leading replica partition function, one has only one phase transition temperature, i.e., $T^{(1)}_{c}$.

\subsection{Finite size effects in the replica field theory} \label{finitesizetemperature}

Here we are investigating temperature effects in a disordered $\lambda\varphi^{4}$ model defined in a $d$-dimensional Euclidean space going beyond the tree-level approximation. We assume that the fields in the replica field theory satisfy periodic boundary condition in Euclidean time and that $k_{c}<1$, where we have spontaneous symmetry breaking.

Periodic boundary condition in Euclidean time implies that this replica field theory is defined in $S^{1}\times\,\mathbb{R}^{d-1}$ with the Euclidean topology for a field theory at finite temperature \cite{Landsman,kapusta,JZinn}. We consider the system  defined in a space with periodic boundary conditions in Euclidean time  using the following non-trivial replica structure given by Eq. (\ref{RSB2}). In this situation, the momentum-space integrals over one component is replaced by a sum over discrete frequencies. Let us define the radius of the compactified dimension of the system by $\beta=T^{-1}$, where $T$ is the temperature of the system.

Let us calculate the one-loop correction to renormalized mass. We have two types of loop-corrections, one from the $\phi^{4}$ vertex, which is written as

\bea
\hspace{-7mm}
[G^{(4)}]_{lm}(x-y,\beta)&=&\sum_{i=1}^{N}\int\,d^{d}z\,[G_{0}]_{li}(x-z,\beta)
\nn\\
&\times&[G_{0}]_{ii}(z-z,\beta)[G_{0}]_{im}(z-y,\beta),\label{sf}
\eea

\noindent and, another contribution, from two $\phi^{3}$ vertices

\begin{align}
\hspace{-3.5mm}
[G^{(3)}]_{lm}(x-y,\beta)&=\sum_{ij=1}^{N}\int\,d^{d}z\,\int\,d^{d}z'\,[G_{0}]_{li}(x-z,\beta)
\nn\\
&\times[G_{0}]^2_{ij}(z-z',\beta) [G_{0}]_{jm}(z'-y,\beta).
\label{sf-sunset}
\end{align}

\noindent To compute the renormalized mass, we must study the amputated correlation function in replica space. At the one-loop approximation, defining $M_{0}^{2}=3\sigma N-2m^{2}_{0}$, the renormalized temperature-dependent mass squared can be written as

\begin{equation}
m^{2}_{R}(M_{0},\beta,\sigma)=m^{2}_{1}(M_{0},\beta,\sigma)+m^{2}_{2}(M_{0},\beta,\sigma)
\end{equation}

\noindent where

\begin{equation}
\hspace{-1.5mm}
m^{2}_{1}(M_{0},\beta,\sigma)=M_{0}^{2}+\frac{\lambda_{0}}{2}\sum_{k=1}^{N}\Bigl(f_{1}(M_{0},\beta;k)+f_{2}(M_{0},\beta,\sigma;k)\Bigr)
\label{fimmm2}
\end{equation}

\noindent and

\bea
m^{2}_{2}(M_{0},\beta,\sigma)&=&\sum_{k=1}^{N}\biggl(m^{2}_{a}(M_{0},\beta;k)+m^{2}_{b}(M_{0},\beta,\sigma;k)
\nn\\
&+&m^{2}_{c}(M_{0},\beta,\sigma;k)\biggr).
\eea

\noindent All these quantities are discussed in the Appendix. For a very large $N$ in $d=4$ the temperature dependent renormalized mass squared can be written as

\bea
m_{R}^{2}(\beta) &=& M_{0}^{2}+\frac{\lambda_{0} N}{4\pi^{2}}\biggl[\sqrt{\frac{\pi}{2}}\sum_{n=1}^{\infty}\sqrt{\frac{M_{0}}{(n\beta)^{3}}}e^{-n\beta M_{0}}
\nn\\
&+&\sqrt{\pi}\biggl(\frac{1}{\sqrt{2}}-\frac{1}{\sigma}\biggr)\sum_{n=1}^{\infty}\frac{1}{\sqrt{n\beta M_{0}}}e^{-n\beta M_{0}}\biggr].
\label{eq:renmassdisorder}                                                              
\eea

\noindent For the large-$N$ limit the thermal mass correction in the one-loop approximation is given by Eq.~(\ref{eq:renmassdisorder}). Notice that in the above equation there is a term proportional to $\sigma^{-1}$, a non-perturbative effect produced by the disorder. In turn, for a weak disorder parameter $\sigma\ll 1$ and for sufficiently small temperatures, $\beta \gg 1$, the last term dominates over the second and the third. In this case it is easy to see that there is a specific temperature in which the renormalized mass squared goes to zero. One says that the system of large-$N$ replica fields presents a phase transition at such a critical temperature. 
One way to proceed is to use the gap equation to obtain non-perturbative results.

In the next section we use again the mean-field descriptions for phase transitions. 
We will restrict our attention to the regime of very low temperatures, investigating a different perturbative expansion for the replica field theory.



\vspace{-3mm}


\section{Replica instantons in the large-$N$ approximation}

The aim of this section is to show the presence of instantons (real or complex) in the model at some range of temperatures. At this point, let us introduce an external source $J_{i}(x)$ in replica space linearly coupled with each replica field. From Eqs. (\ref{NZ}) and (\ref{KactionN}), we are able to define the generating functional of all correlation functions for a large-$N$ Euclidean replica field theory as $\mathbb{E}\,[Z^{N}(J)]=\mathcal{Z}(J)$. Hence it is possible to define the generating functional of connected correlation functions and also the generating functional of one-particle irreducible correlations (vertex functions) in the theory. From the effective action it is possible to find the effective potential of this theory. This is a natural tool to investigate the vacuum structure of the field theory.

However, in the following, we are going to discuss a different perturbative expansion.
Let us define $R(x-y)=\sigma\delta^{d}(x-y)$ and at the large-$N$ limit we must have a fixed $f_{0}=\sigma N$ as we discussed before. We write the replica partition function $\mathcal{Z}(J)$ as a functional differential operator acting on a modified replica partition function without the interaction between the replicas that we call $Q_{0}(J)$. This is a good representation for $\mathcal{Z}(J)$ in the weak disorder limit, and also for $m_{0}^{2}<\sigma N$. The representation for the replica partition function, in the presence of an external source, is similar to the strong-coupling expansion in field theory.  We have

\begin{align}
\hspace{-0.5mm}
\mathcal{Z}(J)&=\exp\biggr[-\frac{1}{2}\sum_{i,j=1}^{N}\int d^{\,d}x\,d^{\,d}y\frac{\delta}{\delta J_{i}(x)}R(x-y)\frac{\delta}{\delta J_{j}(y)}\biggr]
\nn\\
&\times Q_{0}(J),
\label{instanton1}
\end{align}

\noindent where $Q_{0}(J)$ is given by

\begin{equation}
Q_{0}(J)=\int\prod_{j=1}^{N}[d\phi_{j}]\,\exp\biggl(-S_{\textrm{eff}}^{(0)}(\phi_{j},J)\biggr).
\label{instanton2}
\end{equation}

\noindent In the above equation, taking the large-$N$ limit, $S_{\textrm{eff}}^{(0)}(\phi_{i},J)$ is defined as

%


%


%

\begin{align}
\hspace{-0.5mm}
S_{\textrm{eff}}^{(0)}(\phi_{i},J)&=\sum_{i=1}^{N}\int d^{d}x\biggl[\,\,\frac{1}{2}\phi_{i}(x)\Bigl(-\Delta+3f_{0}-2m_{0}^{2}\Bigr)\phi_{i}(x)
\nonumber\\
&+\Bigl(\frac{f_{0}g_{0}}{3!N}\Bigr)^{\frac{1}{2}}\left(1-\frac{m_{0}^{2}}{f_{0}}\right)^{\frac{1}{2}}\phi_{i}^{3}(x) \nonumber \\
 &+\frac{g_{0}}{4!N}\phi_{i}^{4}(x)+J_{i}(x)\phi_{i}(x)\biggr].
\label{instantonN}
\end{align}

The action defined by the above equation describes a large-$N$ replica field theory with two fixed parameters $g_{0}$ and $f_{0}$.  
Notice that all the ultraviolet divergences of this model are fixed by Eqs. (\ref{instanton2}) and (\ref{instantonN}). It is possible to go beyond the tree-level approximation. Working with the bare quantities, and introducing the renormalization constants $Z_{\phi}$, $Z_{g}$ and $Z_{m}$ one is able to renormalize the model for $d\leq 4$. This is the standard procedure. All the divergences of this theory can be eliminated by a wave function, coupling constant and mass renormalization. 
In practice, performing the perturbative expansion defined by Eq. (\ref{instanton1}) is not difficult. For instance, the two-point correlation function is defined as
\begin{equation}
\left. \langle\phi_{i}(x)\phi_{j}(y)\rangle=\frac{\delta^{2}\mathcal{Z}(J)}{\delta J_{i}(x)\delta J_{j}(y)}\right|_{J_{i}=J_{j}=0}.
\end{equation}
\noindent In the following we are interested to go in another direction. We would like to investigate the vacuum structure in the first term of the Eq. (\ref{instanton1}). For each replica field, we can define the following potential $U(\phi)$

\begin{equation}
U(\phi)=\frac{1}{2}(3f_{0}-2m_{0}^{2})\phi^{2}+\frac{\lambda_{0} v}{3!}\phi^{3}+\frac{\lambda_{0}}{4!}\phi^{4},
\label{potential}
\end{equation}

\noindent where $v=\sqrt{6(f_{0}-m_{0}^{2})/\lambda_{0}}$. The false and the true vacuum states $\phi_{\pm}$ are given by

\begin{equation}
\phi_{\pm}=-\frac{3v}{2}\pm3\sqrt{-\frac{f_{0}}{2\lambda_{0}}-\frac{m^{2}_{0}}{6\lambda_{0}}}.\label{raises}
\end{equation}

\noindent Therefore, we obtained the following interesting result: there are instantons in our model. For $f_{0}>m_{0}^{2}>-3f_{0}$, the system develops a spontaneous symmetry breaking in the replica partition function. In this case, all $N$ instantons are complex. On the other hand, for $m_{0}^{2}<-3f_{0}$ we get a similar situation as before, however all the $N$ instantons are 
real \cite{ZinnJustin}.

Let us briefly discuss the decay rate for one replica field in this case of real instantons. Since we would like to discuss such problems exactly as in the bounce problem in quantum mechanics let us define an Euclidean time $\tau$ such that $\phi(x)\equiv\phi(\tau,\vec{x})$. We have a false vacuum in the infinite past and we come back to it in the infinite future

\begin{equation}
\phi(\tau,\vec{x})\rightarrow\phi_{+}, \,\,\,\,\,\,\,\,\tau\rightarrow\pm\infty.
\label{boundary1}
\end{equation}

\noindent In order to have a finite action for the bounce, we also need to go to the vacuum value at spatial infinity. Hence we have

\begin{equation}
\phi(\tau,\vec{x})\rightarrow\phi_{+}, \,\,\,\,\,\,\,\,|\vec{x}|\rightarrow\pm\infty.
\label{boundary2}
\end{equation}

\noindent As discussed in the literature the picture is a formation of bubbles in the middle of the false vacuum. Actually,  asymptotically in Euclidean space the replica configuration is in the false vacuum. A different state appears in the core of the bubble. The probability of decay can be calculated. There is a standard  procedure to find the decay rate in a scalar theory \cite{SColeman1, SColeman2,Flores}. 
One interesting unsolved problem is the phase diagram of liquids in a random porous media. Assuming strong coupling between the fluid and the porous media, for such confined fluids, the random field Ising model is used to describe such systems. These systems can develop a second or a first-order phase transition. Since bubble nucleation is a first order phase transition, we expect that our approach reveals a route to investigate such systems.


\section{Conclusions}\label{conclude}

In this paper we consider a disorder field linearly coupled with the scalar field of the  $\lambda\varphi^{4}$ model defined in a $d$-dimensional Euclidean space.
In the presence of the disorder field, ground state configurations of the field $\varphi(x)$ are defined by a saddle-point equation, where the solutions of such equation depend on particular configurations of the disorder field. As discussed in the literature, perturbation theory is inappropriate to be used in systems where the disorder defines a large number of local minima in the model. One way to circumvent this problem is to average the free energy over the disorder field. 

Recently an alternative approach to obtain the average free energy of this system still using the replicas was proposed. The dominant contribution to the average free energy of this system is written as a series of the integer moments of the partition function of the model. 
Each term of the series defines a replica field theory. 
A crucial point is that, in each replica partition function, all the replica fields must be equal in principle and the number of replica fields must be very large. This shows that we are in the large-$N$ scenario. 
Since we study fluctuations around the saddle-point equations, we obtain two groups. In one group a generic replica partition function with $k\leq k_{c}$, the replica fields are fluctuating around the zero value which is a stable equilibrium state. On the other hand, in the other group of replica partition functions, the zero value of the fields does not describe stable equilibrium states. For replica partitions functions such that $k>k_{c}$ and we must define shifted fields. We establish a connection between spontaneous symmetry breaking mechanism and the structure of the replica space in the disordered model. This was done using a replica symmetry ansatz, the only choice that is consistent with the method. This leads to the aforementioned large-$N$ expansion in Euclidean replica field theory. By investigating finite-size effects in the one-loop approximation, we showed that there is a critical temperature $\beta_{c}^{-1}$ where the renormalized mass is zero.

Also, following the Landau-Ginzburg approach, we obtained that in the case where $m_{0}^{2}\geq N\sigma$, all $N$ replica fields in each replica partition function oscillates around the trivial vacuum. In the large-$N$ approximation, the symmetry $[\mathbb{Z}_{2}\times\mathbb{Z}_{2}\cdots\times\mathbb{Z}_{2}]$ is realized. This range is equivalent to the $\mathbb{Z}_{2}$-symmetric phase for systems without disorder. For $\sigma N > m_{0}^{2}\geq\sigma$, the average free energy is defined by two kinds of replica partition functions with replica fields $\varphi_{i}$ and $\phi_{i}$ respectively. In the large-$N$ limit, as only the leading term is considered, one has that all the replica fields of this leading replica partition function are oscillating around the non-trivial vacuum. For $m_{0}^{2} < \sigma$, all the $N$ replica field in each replica partition function are oscillating around the non-trivial vacuum. Again taking only the leading order term of the series that represent the average free energy, i.e., the large-$N$, the symmetry $[\mathbb{Z}_{2}\times\mathbb{Z}_{2}\cdots\times\mathbb{Z}_{2}]$ is broken. This situation is equivalent to the $\mathbb{Z}_{2}$-broken symmetry for a system without disorder.

Moreover, in the large-$N$ limit, for $m_{0}^{2}<\sigma N$, we wrote the dominant replica partition function as a functional differential operator acting on a modified replica partition function without the interaction between the replicas, which has its similarities with the strong-coupling expansion in field theory. Furthermore, from Eq. (\ref{raises}), the value $m_{0}^{2}=-3\sigma N$ is a boundary between real and complex instantons. For $m_{0}^{2}>-3\sigma N$ there are $N$ complex instantons in the system. For $m_{0}^{2}<-3\sigma N$ the system presents $N$ real instantons. This conclusion is obtained in the diluted instanton approximation. The consequences of these results deserve further investigation.  

A natural continuation of this paper is, for real instantons, to study the system beyond the diluted instanton approximation. 
Another continuation of this paper is to calculate the critical exponents associated with the random field Ising model using the Landau-Ginzburg approach. These issues are under investigation by the authors.


\section*{Acknowlegements}

The authors would like to thank A. Hernandez, G. Krein, E. Curado, S. Alves Dias and T. Micklitz for useful discussions. This paper was partially supported by Conselho Nacional de Desenvolvimento 
Cient\'ifico e Tecnol{\'o}gico (CNPq, Brazil).

\appendix


\section{The temperature-dependent renormalized mass in one-loop approximation}\label{app}

The aim of this Appendix is to discuss the temperature-dependent renormalized mass in one-loop approximation. We consider the system at finite temperature, i.e., with periodic boundary conditions in Euclidean time  using the non-trivial structure in the replica space given by Eq. (\ref{RSB2}).In this situation,  the momentum-space integrals over one component is replaced by a sum over discrete frequencies. For the case of Bose fields we must perform the replacement

\begin{equation}
\int\frac{d^{d}p}{(2\pi)^{d}}f(p)\rightarrow\,\frac{1}{\beta}\int\frac{d^{d-1}p}{(2\pi)^{d-1}}
\sum_{n=-{\infty}}^{\infty}f\biggl(\frac{2n\pi}{\beta},{\bf p}\,\biggr),
\label{aa19}
\end{equation}

\noindent where $\beta$ is the radius of the compactified dimension of the system. This field theory on $S^{1}\times\,\mathbb{R}^{d-1}$ has the Euclidean topology of a field theory at finite temperature. From the two-point Schwinger function we have to calculate

\begin{equation}
\hspace{-1mm}
f_{ij}^{(1)}(M_{0},\beta)=\frac{\delta_{ij}}{\beta}\int\frac{d^{d-1}p}{(2\pi)^{d-1}}
\sum_{n=-\infty}^{\infty}\frac{1}{\bigl((\frac{2\pi n}{\beta})^{2}+{\bf p}^{2}+M_{0}^{2}\bigr)}
\label{aa21}
\end{equation}

\noindent and

\begin{align}
\hspace{-3mm}
f^{(2)}(M_{0},\beta,\sigma)&=\frac{\sigma}{\beta}\int\frac{d^{d-1}p}{(2\pi)^{d-1}}
\sum_{n=-\infty}^{\infty}\frac{1}{\bigl((\frac{2\pi n}{\beta})^{2}+{\bf p}^{2}+M_{0}^{2}\bigr)}
\nonumber\\
&\times\frac{1}{\bigl((\frac{2\pi n}{\beta})^{2}+{\bf p}^{2}+M_{0}^{2}-k\sigma\bigr)},
\label{aa22}
\end{align}

\noindent where ${\bf p}=(p^{2},p^{3},..,p^{d})$.The integral $f_{ij}^{(1)}(M_{0},\beta)$ can be calculated using dimensional regularization \cite{analit1,analit2,analit3,analit4,analit5}. We obtain

\begin{align}
f_{ij}^{(1)}(M_{0},\beta)&=\frac{\delta_{ij}}{2\beta}\frac{1}{(2\sqrt{\pi})^{d-1}}
\Gamma\biggl(\frac{3-d}{2}\biggr)
\nn\\
&\times\sum_{n=-\infty}^{\infty}\frac{1}
{\biggl((\frac{2\pi n}{\beta})^{2}+M_{0}^{2}\biggr)^\frac{3-d}{2}}.
\label{aa23}
\end{align}

\noindent After using dimensional regularization, we have to analytically extend the modified Epstein zeta function \cite{eli,ford}. This zeta function is defined as

\begin{equation}
E(s,a)=\sum_{n=-\infty}^{\infty}\frac{1}{(n^{2}+a^{2})^{s}},
\label{epst}
\end{equation}

\noindent which converges absolutely and uniformly for $\Re(s)>1/2$. Its analytic continuation defines a meromorphic function of $s$ with poles at $s=1/2,-1/2,-3/2,-5/2,...$ and analytic at $s=0$. A useful representation of the analytic extension of this function is

\begin{align}
E(s,a)&= \frac{\sqrt{\pi}}{\Gamma(s)a^{2s-1}}
\nn\\
&\times\biggl[\Gamma\biggl(s-\frac{1}{2}\biggr)+4\sum_{n=1}^{\infty}(n\pi a)^{s-\frac{1}{2}}K_{s-\frac{1}{2}}(2\pi na)\biggr],\label{epst2}
\end{align}

\noindent where $K_{\nu}(z)$ is the modified Bessel function of second kind. Using a modified minimal subtraction renormalization scheme, we discuss each term that contributes to the renormalized mass squared. Using that $f_{1}(M_{0},\beta;k)=\delta^{ij}f_{ij}^{(1)}(M_{0},\beta;k)$ we get

\begin{equation}
\hspace{-3.7mm}
f_{1}(M_{0},\beta;k)=\frac{k}{(2\pi)^{d/2}}\sum_{n=1}^{\infty}\biggl(\frac{M_{0}}{n\beta}\biggr)^{\frac{d}{2}-1}K_{\frac{d}{2}-1}(n\beta M_{0}).
\label{fim}
\end{equation}

Let us discuss $f^{(2)}(\sigma,M_{0},\beta;k)$. We have

\begin{align}
&f^{(2)}(M_{0},\beta,\sigma;k)
\nn\\
&=\frac{\sigma}{\beta}r(d)
\int\,dq\,q^{d-2}\sum_{n=-\infty}^{\infty}
\nn\\
&\times\frac{1}{\bigl((\frac{2\pi n}{\beta})^{2}+q^{2}+M_{0}^{2}\bigr)
\bigl((\frac{2\pi n}{\beta})^{2}+q^{2}+M_{0}^{2}-k\sigma\bigr)},
\label{aa22r}
\end{align}

\noindent where 

\[r(d)=\frac{2\pi^{(d-2)/2}}{\Gamma(\frac{d-2}{2})}\] 

\noindent is an analytic function in $d$. Let us use the following integral

\begin{align}
\hspace{-3mm}
\int_{0}^{\infty}dx\frac{x^{\mu-1}}{(x^{2}+\alpha)(x^{2}+\gamma)}=\frac{\pi}{2}
\frac{\gamma^{\frac{\mu}{2}-1}-\alpha^{\frac{\mu}{2}-1}}{\alpha-\gamma}\csc\left(\frac{\pi\mu}{2}\right).
\label{aa22rr}
\end{align}

\noindent Defining 

\[q(d)=\frac{\pi}{2}r(d)\csc\left[\frac{\pi}{2}(d-1)\right]\] 

\noindent we can write $f^{(2)}(\sigma,M_{0},\beta,k)$ as

\begin{equation}
\hspace{-2mm}
f^{(2)}(M_{0},\beta,\sigma;k)=f^{(21)}(M_{0},\beta;k)+f^{(22)}(M_{0},\beta,\sigma;k),
\end{equation}

\noindent where

\begin{equation}
\hspace{-2mm}
f^{(21)}(M_{0},\beta;k)=-\frac{q(d)}{\beta k}\sum_{n=-\infty}^{\infty}\frac{1}{\bigl((\frac{2\pi n}{\beta})^{2}+M_{0}^{2}\bigr)^{\frac{3-d}{2}}}
\label{aa22rrr}
\end{equation}

\noindent and

\begin{equation}
f^{(22)}(M_{0},\beta,\sigma;k)
=\frac{q(d)}{\beta k}\sum_{n=-\infty}^{\infty}\frac{1}{\bigl((\frac{2\pi n}{\beta})^{2}+M_{0}^{2} -k\sigma\bigr)^{\frac{3-d}{2}}}.
\label{aa22rrr}
\end{equation}

\noindent Using the definition of the Epstein zeta function defined before and $c(d)$ given by

\begin{equation}
c(d)=\frac{\pi^{(3d-6)/2}}{2^{(3-d)}\Gamma(\frac{d-2}{2})}\csc\biggl[\frac{\pi}{2}(d-1)\biggr],
\end{equation}

\noindent we can write $f^{(21)}(M_{0},\beta;k)$ and $f^{(22)}(\sigma,M_{0},\beta,k)$ respectively as

\begin{equation}
f^{(21)}(M_{0},\beta;k)=-\frac{c(d)}{k}\,\beta^{2-d}E\biggl(\frac{3-d}{2},\frac{M_{0}\beta}{2\pi}\biggr)
\label{fim}
\end{equation}

\noindent and

\begin{equation}
\hspace{-1mm}
f^{(22)}(M_{0},\beta,\sigma;k)=\frac{c(d)}{k}\,\beta^{2-d}E\biggl(\frac{3-d}{2},\frac{\beta}{2\pi}\sqrt{M_{0}^{2}-k\sigma}\biggr).
\label{fimm}
\end{equation}

\noindent We will use once again the analytic representation of the function $E(s,a)$ and the modified minimal subtraction renormalization scheme. Defining 

\[g_{1}(M_{0};d,k)=2\pi^{\frac{d-3}{2}}\frac{\Gamma(\frac{d-1}{2})}{\Gamma(\frac{d-2}{2})}\frac{M_{0}^{d-2}}{k}\] 

\noindent and 

\[g_{2}(M_{0};d,k)=2\pi^{\frac{d-1}{2}}\frac{\Gamma(\frac{d-1}{2})}{\Gamma(\frac{d-2}{2})}\frac{(M_{0}^{2}-k\sigma)^{\frac{d-2}{2}}}{k}\]

\noindent we can write the Eq. (\ref{fim}) and Eq. (\ref{fimm}) as

\begin{equation}
\hspace{-2.5mm}
f^{(21)}(M_{0},\beta;k)=-g_{1}\sum_{n=0}^{\infty}(nM_{0}\beta)^{\frac{3-d}{2}}K_{\frac{3-d}{2}}(n\beta M_{0})
\label{f21}
\end{equation}

\noindent and

\begin{align}
f^{(22)}(M_{0},\beta,\sigma;k)&=g_{2}\sum_{n=0}^{\infty}\biggl(n\beta\sqrt{M_{0}^{2}-k\sigma}\biggr)^{\frac{3-d}{2}}
\nn\\
&\times K_{\frac{3-d}{2}}\biggl(n\beta\sqrt{M_{0}^{2}-k\sigma}\biggr).
\label{f22}
\end{align}

\noindent The renormalized temperature-dependent mass squared $m^{2}_{1}(M_{0},\beta,\sigma;k)$ can be written as

\begin{align}
m^{2}_{1}(M_{0},\beta,\sigma;k)&=M^{2}_{0}+\lambda\sum_{k=1}^{N}
\biggl(f_{1}(M_{0},\beta;k)
\nn\\
&+f^{(21)}(M_{0},\beta;k)+f^{(22)}(M_{0},\beta,\sigma;k)\biggr).
\label{fimmm}
\end{align}

\noindent Let us defined $m^{2}_{a}(M_{0},\beta;k)$, $m^{2}_{b}(M_{0},\beta,\sigma;k)$ and $m^{2}_{c}(M_{0},\beta,\sigma;k)$ such that the contribution given by $m_{2}^{2}(M_{0},\beta,\sigma;k)$ is written as

\begin{align}
m^{2}_{2}(M_{0},\beta,\sigma)&=\sum_{k=1}^{N}\biggl(m^{2}_{a}(M_{0},\beta;k)+m^{2}_{b}(M_{0},\beta,\sigma;k)
\nn\\
&+m^{2}_{c}(M_{0},\beta,\sigma;k)\biggr).
\end{align}

\noindent We have

\begin{equation}
\hspace{-1mm}
m^{2}_{a}(M_{0},\beta;k)=\frac{k}{\beta}\int\frac{d^{d-1}p}{(2\pi)^{d-1}}
\sum_{n=-\infty}^{\infty}\frac{1}{\bigl((\frac{2\pi n}{\beta})^{2}+{\bf p}^{2}+M_{0}^{2}\bigr)^{2}},
\end{equation}

\begin{align}
&m^{2}_{b}(M_{0},\beta,\sigma;k)
=\frac{k\sigma}{\beta}\int\,\frac{d^{d-1}p}{(2\pi)^{d-1}}
\sum_{n=-\infty}^{\infty}
\nn\\
&\times
\frac{1}{\bigl((\frac{2\pi n}{\beta})^{2}+{\bf p}^{2}+M_{0}^{2}\bigr)^{2}\bigl((\frac{2\pi n}{\beta})^{2}+{\bf p}^{2}+M_{0}^{2}-k\sigma\bigr)}
\end{align}

\noindent and finally

\begin{align}
&m^{2}_{c}(M_{0},\beta,\sigma;k)
=\frac{\sigma^{2}}{\beta}\int\,\frac{d^{d-1}p}{(2\pi)^{d-1}}
\sum_{n=-\infty}^{\infty}
\nn\\
&\times
\frac{1}{\bigl((\frac{2\pi n}{\beta})^{2}+{\bf p}^{2}+M_{0}^{2}\bigr)^{2}
\bigl((\frac{2\pi n}{\beta})^{2}+{\bf p}^{2}+M_{0}^{2}-k\sigma\bigr)^{2}}.
\end{align}

\noindent After using dimensional regularization and considering the analytical extension for the Epstein function we can write $m^{2}_{a}(M_{0},L;k)$ as

\begin{equation}
m^{2}_{a}(M_{0},\beta;k)=\frac{M_{0}^{d-4}k}{(2\pi)^{d/2}}\sum_{n=1}^{\infty}(n\beta M_{0})^{\frac{4-d}{2}}K_{\frac{4-d}{2}}(n\beta M_{0}).
\end{equation}

\noindent To solve the integral in $m^{2}_{b}$ and $m^{2}_{c}$ we can use the Feynman parametrization

\begin{equation}
\frac{1}{a^{s}b^{l}}=\frac{\Gamma(s+l)}{\Gamma(s)\Gamma(l)}\int_{0}^{1}dx\,\frac{x^{s-1}(1-x)^{l-1}}{[ax+b(1-x)]^{s+l}},
\end{equation}

\noindent to write the respective integrands in an adequate form. After this and using the expression

\begin{widetext}

\begin{equation}
\int\frac{d^{d}q}{(2\pi)^{d}}\frac{(q^{2})^{a}}{(q^{2}+A)^{b}}=\frac{\Gamma(b-a-d/2)\Gamma(a+d/2)}{(4\pi)^{\frac{d}{2}}\Gamma(b)\Gamma(d/2)}
A^{-(b-a-d/2)},
\end{equation}

\noindent and defining 

\vspace{-0.5cm}

$$h_{1}(d)=\frac{1}{(2\pi)^{\frac{d}{2}}(d-3)(d-5)}\Gamma\biggl(\frac{7-d}{2}\biggr)$$

\noindent and 

\vspace{-0.5cm}

$$h_{2}(d)=\frac{1}{(2\pi)^{\frac{d}{2}}(d-3)(d-5)(d-7)}\Gamma\biggl(\frac{9-d}{2}\biggr)$$

\noindent such contributions are given by

\begin{align}
m^{2}_{b}(M_{0},\beta,\sigma;k)=&h_{1}(d)\biggl[
\frac{\sqrt{8}(d-3)}{\Gamma\bigl(\frac{5-d}{2}\bigr)}\frac{kM_{0}^{d-4}}{\sigma}\sum_{n=1}^{\infty}
\bigl(n\beta M_{0}\bigr)^{\frac{4-d}{2}}K_{\frac{4-d}{2}}(n\beta M_{0})\\  \nonumber
&-\frac{16}{\Gamma\bigl(\frac{3-d}{2}\bigr)}\frac{M_{0}^{d-2}}{\sigma^{2}}\sum_{n=1}^{\infty}
\bigl(n\beta M_{0}\bigr)^{\frac{2-d}{2}}K_{\frac{2-d}{2}}(n\beta M_{0})\\ \nonumber
&+\frac{16} {\Gamma\bigl(\frac{3-d}{2}\bigr)}\frac{(M_{0}^{2}-k\sigma)^\frac{{d-2}}{2}}{\sigma^{2}}\sum_{n=1}^{\infty}
\biggl(n\beta \sqrt{M^{2}_{0}-k\sigma}\biggr)^{\frac{2-d}{2}}
K_{\frac{2-d}{2}}\biggr(n\beta\sqrt{M^{2}_{0}-k\sigma}\biggl)\biggr]
\end{align}

\noindent and

\begin{align}
m^{2}_{c}(M_{0},\beta,\sigma;k)=&h_{2}(d)\biggl[
\frac{\sqrt{32} (d-3)}{\Gamma\bigl(\frac{5-d}{2}\bigr)}\frac{M_{0}^{d-4}}{k^{2}}\sum_{n=1}^{\infty}
\bigl(n\beta M_{0}\bigr)^{\frac{4-d}{2}}K_{\frac{4-d}{2}}(n\beta M_{0})\\ \nonumber
&-\frac{32}{\Gamma\bigl(\frac{3-d}{2}\bigr)}\frac{M_{0}^{d-2}}{k^{3}\sigma}\sum_{n=1}^{\infty}
\bigl(n\beta M_{0}\bigr)^{\frac{2-d}{2}}K_{\frac{2-d}{2}}(n\beta M_{0}) 
+\frac{2(d-3)}{\Gamma\bigl(\frac{7-d}{2}\bigr)}\frac{M_{0}^{d-6}}{k^{2}}\sum_{n=1}^{\infty}
\bigl(n\beta M_{0}\bigr)^{\frac{6-d}{2}}K_{\frac{6-d}{2}}(n\beta M_{0})\\ \nonumber  
&+\frac{32}{\Gamma\bigl(\frac{3-d}{2}\bigr)}\frac{(M_{0}^{2}-k\sigma)^{\frac{d-2}{2}}}{k^{3}\sigma}\sum_{n=1}^{\infty}
\biggl(n\beta\sqrt{M_{0}^{2}-k\sigma}\biggr)^{\frac{2-d}{2}}K_{\frac{2-d}{2}}\biggl(n\beta\sqrt{M_{0}^{2}-k\sigma}\biggr)\biggr].
\end{align}

\end{widetext}


\end{document}